# Conformational landscapes in cryo-ET data based on MD simulations

Slavica Jonic*

IMPMC-UMR 7590 CNRS, Sorbonne Université, MNHN, Paris, France

*corresponding author

**Contact details of the corresponding author**:

Slavica Jonic

IMPMC - UMR 7590 CNRS

Sorbonne Université, CC 115

4 place Jussieu, 75005 Paris, France

Phone: +33 1 44 27 72 05

Fax: +33 1 44 27 37 85

E-mail: slavica.jonic@sorbonne-universite.fr

**Abstract:** Cryo-electron tomography (cryo-ET) provides a unique window into molecular organization in cellular environments (*in situ*). However, the interpretation of molecular structural information is complicated by several intrinsic properties of cryo-ET data, such as noise, missing wedge, and continuous conformational variability of the molecules. Additionally, in crowded *in situ* environments, the number of particles extracted is sometimes small and precludes extensive classification into discrete states. These challenges shift the emphasis from high-resolution structure determination toward validation and interpretation of low-resolution density maps, and analysis of conformational flexibility. Molecular Dynamics (MD) simulations are particularly well suited to this task, as they provide a physically grounded way to explore continuous conformation transitions consistent with both experimental data and molecular energetics. This review focuses on the roles of MD simulations in cryo-ET, emphasizing their use in emerging methods for conformational landscape determination and their contribution to gain new biological insight.

## 1. Introduction

Cryo-electron tomography (cryo-ET) provides a unique window into molecular organization *in situ* [1-6]. While operating in low electron dose and cryogenic conditions, it enables three-dimensional (3D) visualization of biomolecular complexes in their native cellular environments. However, the crowded and heterogeneous cellular contexts and low signal-to-noise ratios (SNRs) of cryo-ET data complicate the analysis of structural and conformational variability of individual molecular complexes [7-12]. Additionally, 3D reconstructions (tomograms) from collected cryo-ET images (tilt series) are affected by deformations that result from a limited angular range of the collected data, which is known as missing wedge (MW). This further complicates the determination of the structure of individual molecular complexes from the data.

As a consequence, cryo-ET data analysis workflows for structure determination usually rely on methods for sorting data of individual molecular complexes into discrete classes and class averaging [13-19]. However, discrete classifications are naturally suited to discrete well-separated structural states and hinder revealing subtle structural differences between individual complexes that can be hidden in class averages. Additionally, a small number of particles extracted from *in situ* cryo-ET data in some cases precludes extensive classification into discrete states.

On the other hand, tomographic analyses increasingly reveal ensembles of related conformations rather than discrete structural states [7,12,20-22]. This challenges traditional structural biology paradigms and motivates the development of computational approaches capable of interpreting cryo-ET data in terms of conformational landscapes rather than static models. Recently, a few such new methods have been proposed [23-26]. They analyze conformations of individual molecular instances in the data instead of using discrete classifications.

Therefore, a defining feature of current structural studies by cryo-ET is not merely limited structural resolution due to intrinsic properties of cryo-ET data but also due to the presence of gradual conformational transitions with many intermediate states, known as continuous conformational heterogeneity. Molecular dynamics (MD) simulations and related mechanics-based modeling approaches provide a natural framework to address this challenge. MD simulations encode physical atomic interactions and trajectories in conformational space, and offer a natural way to relate experimental density to molecular motion. Over the last decade, MD-based methods have evolved from validating and interpretive tools, applied post hoc to experimentally derived structures, into integrated components of cryo-ET analysis pipelines, capable of sampling conformational landscapes under experimental constraints.

This review focuses on the roles of MD simulations in cryo-ET, emphasizing their use in emerging methods for conformational landscape determination and how these approaches enable new biological insight. Particular attention is paid to conceptual and historical links between single-particle cryo-electron microscopy (cryo-EM) and cryo-ET conformational variability methods, and to recent cryo-ET studies that motivate future methodological developments for determination of conformational landscapes.

## 2. Conformational landscapes in cryo-ET data based on MD simulations

*Conformational landscape representations: space of atomic or coarse-grained models vs. space of density maps*

Recent deep learning approaches for conformational heterogeneity analysis in cryo-ET, such as tomoDRGN [25], cryoDRGN-ET [26], and cryoDRGN-AI [27], highlight the growing interest in conformational landscape reconstruction from tomographic data. These methods infer low-dimensional conformational landscapes from 2D particle tilt images, enabling visualization of continuous conformational variability without requiring prior atomic models. While powerful for uncovering heterogeneity in cryo-ET data, such approaches yield conformational landscapes expressed in terms of particle density-map predictions, rather than explicit atomic models or trajectories. In these methods, the conformational landscape is a low-dimensional latent space into which high-dimensional input data are embedded by deep learning. In contrast, hybrid methods based on dynamics simulations, such as MDTOMO [24] and HEMNMA-3D [23], yield conformational landscapes expressed in terms of atomic or coarse-grained models. Among them, MDTOMO integrates classical MD simulations (based on Newton's equations of motion) with cryo-ET data, providing a direct link between experimentally observed heterogeneity and physically grounded models of conformational transitions.

The current methods for determination of conformational landscapes in cryo-ET data are summarized in **Table 1**.

*Integration of classical MD simulations with cryo-ET data: Conceptual bridges with single-particle cryo-EM*

Early cryo-EM MD-based approaches demonstrated how cryo-EM density maps could bias MD trajectories to refine atomic models [28,29]. These methods focused on deriving atomic models from consensus maps and did not address particle-to-particle variability. A major conceptual advance came from methods that treated continuous conformational heterogeneity in 2D single particle images explicitly. HEMNMA and its deep learning extension DeepHEMNMA introduced analysis of continuous variability in 2D cryo-EM images based on flexible fitting of particle images using normal modes [30,31]. Normal modes approximate the full dynamics by a linear combination of harmonic motions and are a fast alternative to full MD simulations. MDSPACE extended this per-particle flexible fitting framework to full MD-driven conformational landscape inference guided by cryo-EM data [32]. In MDSPACE, MD simulations are directly constrained by experimental images, producing physically realistic atomic-scale conformational landscapes.

HEMNMA-3D [23] and MDTOMO [24] extend these paradigms to cryo-ET data analysis. MDTOMO is a cryo-ET extension of MDSPACE: both perform full per-particle MD-based flexible fitting and landscape inference, but differ in whether experimental constraints arise from 2D cryo-EM single particle images (MDSPACE) or 3D subtomograms (MDTOMO). Similarly, HEMNMA-3D extends HEMNMA from single-particle cryo-EM to cryo-ET. These MD and normal-mode methods for conformational landscape determination in cryo-EM and cryo-ET are part of ContinuousFlex [33,34], the plugin for Scipion [35].

*Classical MD simulations vs. normal modes for determination of conformational landscapes in cryo-ET data*

Currently, the finest conformational landscapes can be obtained with MDTOMO [24]. In MDTOMO, MD simulations are biased by forces derived from cryo-ET subtomogram densities, enabling flexible fitting of molecular models into individual subtomograms. MDTOMO produces a set of structures that lie on a high-dimensional conformational landscape whose low-dimensional projections reveal relationships between different conformational states and dominant modes of conformational variability.

Specifically, given an initial atomic or coarse-grained model, and a set of subtomograms with initial poses, MDTOMO flexibly fits the model against each subtomogram while performing local pose refinement. Then, the flexibly fitted models are rigid-body aligned and projected onto a low-dimensional space (usually PCA or UMAP space of less than 10 dimensions). This low-dimensional space (conformational landscape) is then analyzed using semi-automatic tools provided in ContinuousFlex, which for instance allow animations of atomic-model motions along selected directions in the space, calculation of subtomogram averages along these directions, or clustering [33].

A complementary approach is provided by HEMNMA-3D, which adapts normal mode analysis to cryo-ET subtomograms [23]. Instead of full MD, HEMNMA-3D relies on elastic network models to describe low-frequency collective motions. Subtomograms are flexibly fitted by varying amplitudes of normal modes, yielding an approximation of the conformational landscape consistent with experimental subtomograms. Although HEMNMA-3D does not include explicit atomic interactions, it shares with MDTOMO the core conceptual feature of flexible fitting of a given conformation into individual subtomograms and continuous landscape inference. Both methods produce conformational landscapes that show relationships between different conformational states, but HEMNMA-3D provides a fast approximation of the actual landscape, requires no prior pose knowledge, and allows to use non-atomic initial models, such as cryo-EM maps or subtomogram averages (in such cases, the initial model is a cloud point representation of the input density map and can be moved with normal modes) [23].

*Uncertainties due to noise*

MDTOMO can deal with uncertainties due to noise during MD-based fitting or after the conformational landscape is obtained (post hoc). During fitting, it can use the enhanced sampling technique of replica-exchange umbrella sampling (REUS) [36]. In ContinuousFlex, REUS is provided as an option for running both MDSPACE and MDTOMO [33]. Using REUS, multiple replicas of the system can be simulated in parallel under different weights of the biasing potential (the data term that defines experimental restraints), with exchanges between replicas enabling efficient exploration of conformational space. REUS mitigates trapping in local minima induced by noise in the data and enables more exhaustive sampling of alternative conformations compatible with the data. As such, enhanced-sampling MD bridges physical modeling and uncertainty quantification within a single framework. This was demonstrated in the case of MD-based flexible fitting of individual cryo-EM density maps [36,37].

When REUS is not used, uncertainties can be treated post hoc. The post hoc approach consists of averaging individual atomic models and the corresponding experimental data instances locally in the resulting conformational landscape. By calculating subtogram averages in local dense regions of the conformational landscape, MDTOMO and HEMNMA-3D fitting errors due to noise can be minimized [23,24].

Therefore, MDTOMO and HEMNMA-3D do not explicitly quantify uncertainties, but can mitigate them efficiently. On the other hand, approaches that focus on quantifying uncertainty have been proposed for cryo-EM [38,39]. While they are not focused on conformational landscape determination, they address a related problem of determining conformational ensembles that satisfy the restraints defined by the data. They differ fundamentally in how structural ensembles are generated, sampled, and interpreted. However, a common characteristic for these methods is that the sampling is performed in an abstract space of all possible molecular conformations (i.e., sets of atomic or coarse-grained coordinates) regardless of physical time, which can be performed using Bayesian frameworks. They result in statistical weights (posterior probabilities) for candidate structures generated externally (e.g., via MD simulations or normal-mode analysis) [38] or in posterior ensembles based on stochastic sampling (Monte-Carlo–like) [39], reflecting agreement with the data. However, they do not encode physical connectivity (dynamics) between states.

In contrast, MDTOMO and its cryo-EM counterpart MDSPACE perform sampling on trajectories with physical connectivity between sampled conformations from an initial state to the state in the given particle data (cryo-ET subtogram or cryo-EM particle image). By analyzing multiple data instances, each of them reconstructs a statistical conformational landscape, where each state reflects a physically plausible conformation consistent with experimental data. Although states in the final landscape are not directly connected by MD trajectories, each simulated particle pathway ensures mechanistic realism, and inter-state relationships can be inferred based on proximity in the landscape. This combination of MD-based connectivity and landscape statistics allows for steered MD or targeted MD simulations between landscape states to probe transition pathways, bridging statistical inference and mechanistic interpretation.

The discussed MD-based and normal-mode-based cryo-EM and cryo-ET computational methods are summarized in **Table 2**.

### 3. From MD simulations as validation and interpretation tools to atomistic conformational landscapes *in situ*

A key distinction between MD-based validation and interpretation cryo-ET studies and emerging MD-based conformation-landscape determination methods for cryo-ET lies in their treatment of structural variability. In the context of validation and interpretation, MD simulations are primarily used post hoc to assess the physical plausibility, stability, or local dynamics of specific structures or interaction geometries observed by cryo-ET. These approaches do not aim to reconstruct continuous conformational variability of individual macromolecules from tomographic data. By contrast, in the conformational-landscape determination context, MD simulations are integrated with cryo-ET data analysis via flexible fitting.

MDTOMO is currently the only method that uses classical MD simulations to reconstruct high quality, atomic-level conformational landscapes from cryo-ET data. HEMNMA-3D uses reduced mechanics simulations, based on normal mode analysis, and represents a fast alternative to MDTOMO.

MDTOMO helped to determine the conformational landscape of SARS-CoV-2 spike glycoproteins *in situ*, moving beyond discrete subtomogram averages [21,40,41] toward a continuous, MD-driven description of motions of receptor-binding and N-terminal domains, and detected variability inaccessible to averaging-based approaches, such as independent motions of the spike's domains [24]. HEMNMA-3D characterized for the first time breathing and gapping motions of nucleosomes *in situ* [23], previously detected via manual measurements *in situ* [20] and theoretically [42,43]. This contrasts with other chromatin studies where MD simulations are applied to multi-nucleosome assemblies and crowded nuclear environments, focusing on the physical plausibility and local dynamics of specific chromatin configurations rather than reconstructing the intrinsic conformational landscape of individual nucleosomes [44-47].

Importantly, many biological systems are natural candidates for extension from post hoc MD-based interpretation toward conformational landscape inference. For viral surface proteins, such as the SARS-CoV-2 spike, landscape-based approaches could extend current studies beyond validation of models and their flexibility to the full reconstruction of continuous conformational variability of the stalk and the receptor-binding domain from tomographic data, enabling quantitative comparisons between viral strains, liganded states, or membrane environments. Also, conformational landscape inference could capture flexibility of bacterial and archaeal surface assemblies, such as competence pili and S-layers; it could shed light on continuous variability associated with pilus extension–retraction cycles [22] and curvature adaptation of S-layer lattices on complex cell surfaces [48,49], rather than validating a single model. In cytoskeletal motor systems such as dynein-2 bound to microtubule doublets, reconstructing conformational landscapes could move the studies beyond a molecular-scale mechanistic interpretation of their binding preference observed by cryo-ET [50] and provide a means to relate *in situ* structural heterogeneity to mechanochemical states along the motor cycle.

Together, these examples illustrate how the methodological advances embodied by MDTOMO, HEMNMA-3D, and related approaches could transform many current applications of MD from validation tools into systematic frameworks for reconstructing conformational landscapes of macromolecular assemblies directly *in situ*.

The discussed examples of cryo-ET studies based on MD simulations or normal mode analysis are summarized in **Table 3**, along with biological insight gained, motivating the development and use of conformational landscape methods.

4. **Outlook and open challenges**

*Future improvements in data quality and computational speed*

In the future, in parallel with continued methodological advances on the modeling and simulation side, further progress in cryo-ET sample preparation, instrumentation, and computational workflows and infrastructure will be essential to fully realize the potential of atomic-scale conformational landscape inference *in situ*.

Improvements in specimen vitrification, focused ion beam milling, and lamella preparation are expected to yield thinner, more uniform samples with reduced artifacts and improved preservation of native cellular architectures. On the instrumentation side, advances in phase plates, direct electron detectors, energy filtering, and optimization of dose-fractionation and tilt-scheme strategies will further enhance SNR of cryo-ET data. Coupled with advances in computational approaches for MW correction, denoising, and particle picking, these developments will increase the fidelity with which subtle conformational heterogeneity can be detected and quantified from cryo-ET data. Continued growth in computational power and simulation efficiency will make it increasingly feasible to apply MD-driven conformational landscape inference to larger systems and longer timescales.

Together, improvements in experimental data quality and computational speed, coupled with methodological advances for modeling and simulation, will transform the current integration of cryo-ET with MD simulations into a broadly applicable framework for reconstructing macromolecular dynamics *in situ*.

*Going beyond conformational landscapes of individual molecular complexes*

Current atomic-scale conformational landscape inference methods remain limited in scope. To date, MDTOMO and HEMNMA-3D have primarily been demonstrated on individual, relatively compact macromolecular complexes, such as viral spikes or nucleosomes, where a single dominant molecular entity can be reasonably segmented from subtomograms and described starting from a single atomic model. Extending these approaches to larger, composite systems, such as multiple interacting nucleosomes within chromatin fibers, arrays of membrane protein complexes embedded in curved membranes, or repeated assemblies within pili, S-layers, or cytoskeletal filaments, poses substantial conceptual and computational challenges. Such systems exhibit collective, multi-body conformational variability, long-range mechanical coupling, and strong geometric constraints imposed by the surrounding cellular environment.

Addressing these challenges would require several methodological advances. Among them are scalable representations of conformational variability (e.g., multiscale modeling that combines local atomic flexibility with coarse-grained collective modes spanning multiple interacting units of an assembly), improved strategies for segmentation and contextual modeling in cryo-ET (e.g., to define parts that should be treated as flexible or quasi rigid), and efficient conformation sampling schemes.

Meeting these challenges would allow atomic-scale conformational landscape inference to move beyond isolated complexes toward mesoscale assemblies *in situ*, enabling reconstruction of conformational dynamics across chromatin fibers, membrane protein networks, and supramolecular machines embedded within native cellular architectures. Such developments would represent a major step toward a truly integrative, physics-based description of structural variability and dynamics in the cellular context, bridging molecular simulations and cryo-ET across scales.

On the other hand, deep learning frameworks can be a more efficient alternative to address the dynamics of such large assemblies in cellular environments. As being able to efficiently embed complex information into a compact, low-dimensional representation, they could potentially be used to embed multi-scale dynamics of large assemblies from large cryo-ET datasets, followed by MD simulations for atomistic interpretation of the embeddings.

**Table 1**: Current methods for determination of conformational landscapes in cryo-ET data.

| Method | Target | Sampling mechanism | Role of MD simulations | Prior pose estimation | Prior structural knowledge | Output |
|---|---|---|---|---|---|---|
| **MDTOMO** [24] | 3D particle subtomograms | Full MD–driven sampling, optional REUS | Central (explicit MD trajectories) | Prior pose estimation required, but local refinement | Atomic or coarse-grained model of one (initial) conformation required | Conformational landscapes expressed in terms of atomic or coarse-grained models (with optional density maps) |
| **HEMNMA-3D** [23] | 3D particle subtomograms | Normal mode–based sampling | Reduced mechanics (normal modes) | No prior pose estimation required | Atomic model, coarse-grained model or density map of one (initial) conformation required | Conformational landscapes expressed in terms of atomic or coarse-grained models (with optional density maps) or in terms of density maps (depending on the type of the initial model) |
| **tomoDRGN** [25] | 2D particle tilt images | Neural latents | Fully data-driven (no MD simulations) | Prior pose estimation required, no refinement | Prior structural knowledge required for pose estimation, but derived from data | Conformational landscapes expressed in terms of density maps |
| **cryoDRGN-ET** [26] | 2D particle tilt images | Neural latents | Fully data-driven (no MD simulations) | Prior pose estimation required, no refinement | Prior structural knowledge required for pose estimation, but derived from data | Conformational landscapes expressed in terms of density maps |
| **cryoDRGN-AI** [27] | 2D particle tilt images | Neural latents | Fully data-driven (no MD simulations) | No prior pose estimation required | No prior structural knowledge required (*ab initio*) | Conformational landscapes expressed in terms of density maps |

**Table 2**: MD-based or normal-mode-based cryo-EM and cryo-ET computational methods discussed in this article.

| Method | Imaging modality | Target | Sampling mechanism | Role of MD simulations | Output |
|---|---|---|---|---|---|
| **MDTOMO** [24] | Cryo-ET | 3D particle subtomograms | Full MD–driven sampling, optional REUS | Central (explicit MD trajectories) | Conformational landscapes |
| **HEMNMA-3D** [23] | Cryo-ET | 3D particle subtomograms | Normal mode–based sampling | Reduced mechanics (normal modes) | Conformational landscapes |
| **MDSPACE** [32] | Cryo-EM | 2D particle images | Full MD–driven sampling, optional REUS | Central (explicit MD trajectories) | Conformational landscapes |
| **HEMNMA/ DeepHEMNMA** [30,31] | Cryo-EM | 2D particle images | Normal mode–based sampling | Reduced mechanics (normal modes) | Conformational landscapes |
| **BioEM** [38] | Cryo-EM | 2D particle images | External sampling of candidate models | MD or normal modes can be used to pre-generate models | Posterior probabilities |
| **Metainference** [39] | Cryo-EM | 3D density maps | MD-driven stochastic sampling (Monte-Carlo–like) | Central (sampling engine for posterior) | Posterior ensembles |

**Table 3**: Examples of cryo-ET studies based on MD simulations or normal mode analysis discussed in this article, with biological insight gained, motivating the development and use of conformational landscape methods.

| System | Method | Insight | Study |
|---|---|---|---|
| **SARS-CoV-2 spike** | Cryo-ET and MD-informed interpretation | Hinge flexibility, stalk bending, and receptor-binding domain mobility *in situ* | Turoňová et al., 2020 [21] |
| | Conformational landscape determination by MDTOMO | Independent motion of each receptor-binding domain and N-terminal domain of the spike | Vuillemot et al., 2023 [24] |
| **Chromatin** | Cryo-ET and MD-informed interpretation | Varying distance between DNA of individual nucleosomes *in situ* was interpreted by different modes of nucleosome opening, among which "gaping" of the particle | Eltsov et al., 2018 [20] |
| | Conformational landscape determination by HEMNMA-3D | Breathing and gaping motions of individual nucleosomes characterized for the first time *in situ* | Harastani et al., 2021 [23] |
| | Cryo-ET and MD-informed interpretation | Mechanistic insight into the physical plausibility and local dynamics of experimentally observed chromatin architectures | Kreysing et al., 2025 [46] |
| **Competence pilus machine** | Cryo-ET and MD-informed validation and interpretation | Pilus extension and retraction explored by MD simulations, revealing gate-opening mechanisms | Maggi e al., 2025 [22] |
| **Archaeal S-layer** | Cryo-ET and MD-informed interpretation | Pseudohexagonal geometry and flexibility of the S-layer and its cation-binding properties explored by MD simulations | von Kügelgen et al., 2024 [49] |

| Dynein-2 bound to ciliary doublet microtubules | Cryo-ET and MD-informed interpretation | Preferential binding of dynein-2 to A-tubule observed by cryo-ET was explained by MD simulations, indicating that interactions with tyrosinated tubulin lattices stabilize binding | He at al., 2025 [50] |

**Acknowledgement**

The author acknowledges the support of the French National Research Agency – ANR (ANR-23-CE45-0012-03 and ANR-25-CE11-5519-01) and access to HPC resources of CINES and IDRIS granted by GENCI (AD010710998R3, AD010714089R1).

In this study, *in-situ* cryo-ET subtomogram averaging and modeling were used to reveal the architecture of the competence pilus machine (CPM) that adopts multiple conformational states. The subtomogram average was used as an envelope to generate a full-length pseudoatomic model of the CPM. All-atom MD simulations were then performed to check structural integrity of this model. Also, steered MD simulations were used to explore pilus extension and retraction, revealing gate-opening mechanisms. Thus, MD simulations are used for validation and interpretation of the results.

This study employs all-atom MD simulations to explore stability and local dynamics of specific chromatin arrangements resolved by cryo-ET at nuclear periphery. Here, MD simulations are initialized from cryo-ET–derived configurations and used to interpret nucleosome packing, linker DNA flexibility, and heterogeneity within crowded nuclear environments, but the simulations are not directly biased by experimental density or population-level statistics. Thus, in this study, MD simulation serves as a structure-informed, post hoc interpretive tool, providing mechanistic insight into the physical plausibility and local dynamics of experimentally observed chromatin architectures.

In this study, all-atom MD simulations combined with cryo-ET allowed to examine cation-binding properties of an archaeal S-layer. The study suggests that the S-layer is pseudohexagonal and supports flexibility, allowing the S-layer protein to coat different parts of the cell surface of varying curvature with near-perfect continuity. Thus, MD simulations inform interpretation in this study.

In this study of dynein-2 motors bound to ciliary doublet microtubules, MD simulations were used to provide a molecular-scale mechanistic interpretation of their binding preference observed by cryo-ET. Cryo-ET revealed that dynein-2 preferentially binds the A-tubule of the microtubule doublet. Coarse-grained MD simulations were subsequently used to examine interactions between the dynein microtubule-binding domain and tubulin, showing that preferential interactions with tyrosinated tubulin lattices stabilize this binding geometry.